\documentclass[10pt,twocolumn]{article}

\usepackage[T1]{fontenc}
\usepackage[utf8]{inputenc}

\usepackage{amsmath}
\usepackage{amssymb}
\usepackage{mathtools}
\usepackage{authblk}
\usepackage{cite}
\usepackage{url}
\usepackage{graphicx}
\usepackage{booktabs}
\usepackage{tabularx}
\usepackage{ragged2e}
\usepackage{stmaryrd}

\newcolumntype{Y}{>{\RaggedRight\arraybackslash}X}

\usepackage{pgfplots}
\pgfplotsset{compat=1.18}
\usepackage{siunitx}
\usepackage{sansmath}
\usepackage{comment}

\usepackage{listings}
\usepackage{xcolor}
\usepackage[scaled=0.92]{inconsolata}

\definecolor{codebg}{HTML}{F8F8F8}
\definecolor{codeborder}{HTML}{DDDDDD}
\definecolor{keyword}{HTML}{0055AA}
\definecolor{comment}{HTML}{778899}
\definecolor{string}{HTML}{AA5500}

\lstdefinestyle{paper}{
  backgroundcolor=\color{codebg},
  frame=single,
  rulecolor=\color{codeborder},
  frameround=tttt,
  numbers=left,
  numberstyle=\tiny\color{comment},
  numbersep=8pt,
  basicstyle=\ttfamily\small,
  keywordstyle=\bfseries\color{keyword},
  commentstyle=\itshape\color{comment},
  stringstyle=\color{string},
  breaklines=true,
  showstringspaces=false,
  tabsize=2,
  captionpos=b,
  upquote=true
}

\newcommand{\M}{\mathcal{M}}
\newcommand{\K}{\mathcal{K}}
\newcommand{\LIR}{\mathcal{L}}

\newcommand{\sem}[1]{\llbracket #1 \rrbracket}

\title{Hexagon-MLIR: An AI Compilation Stack For Qualcomm's Neural Processing Units (NPUs)}

\author[1]{Mohammed Javed Absar}
\author[2]{Muthu Baskaran}
\author[2]{Abhikrant Sharma}
\author[3]{Abhilash Bhandari}
\author[2]{Ankit Aggarwal}
\author[3]{Arun Rangasamy}
\author[3]{Dibyendu Das}
\author[2]{Fateme Hosseini}
\author[2]{Franck Slama}
\author[2]{Iulian Brumar}
\author[2]{Jyotsna Verma}
\author[3]{Krishnaprasad Bindumadhavan}
\author[2]{Mitesh Kothari}
\author[2]{Mohit Gupta}
\author[3]{Ravishankar Kolachana}
\author[2]{Richard Lethin}
\author[2]{Samarth Narang}
\author[3]{Sanjay Motilal Ladwa}
\author[3]{Shalini Jain}
\author[2]{Snigdha Suresh Dalvi}
\author[2]{Tasmia Rahman}
\author[3]{Venkat Rasagna Reddy Komatireddy}
\author[3]{Vivek Vasudevbhai Pandya}
\author[2]{Xiyue Shi}
\author[2]{Zachary Zipper}

\affil[1]{Qualcomm Technologies International, Ltd.}
\affil[2]{Qualcomm Technologies, Inc.}
\affil[3]{Qualcomm India Private Limited}

\date{}

\begin{document}
\maketitle

\begin{abstract}
In this paper, we present Hexagon-MLIR \cite{HexagonMLIRGitHub, HexagonMLIRBlog},
 an open-source compilation stack that targets Qualcomm Hexagon Neural Processing Unit (NPU) \cite{Qualcomm2024NPU}
and provides unified support for lowering Triton kernels \cite{TRITON} and PyTorch models \cite{PyTorch}.
Built using the MLIR framework \cite{mlir}, our compiler applies a structured sequence of passes to exploit 
NPU architectural features to accelerate AI workloads.
It enables faster deployment of new Triton kernels (hand-written or subgraphs from PyTorch 2.0 \cite{Ansel}),
for our target by providing automated compilation from kernel to binary. By ingesting Triton kernels, 
we generate \textit{mega-kernels} that maximize data locality in the NPU's Tightly Coupled Memory (TCM),
reducing the bandwidth bottlenecks inherent in library-based approaches.
This initiative complements our commercial toolchains by providing developers with an open-source MLIR-based compilation stack
that gives them a path to advance AI compilation capabilities through a more flexible approach. 
Hexagon-MLIR is a work-in-progress, and we are continuing to add many more optimizations and capabilities in this effort.
\end{abstract}

%--------------------------------
\begin{figure*}[!t]
\centering
\includegraphics[width=\linewidth]{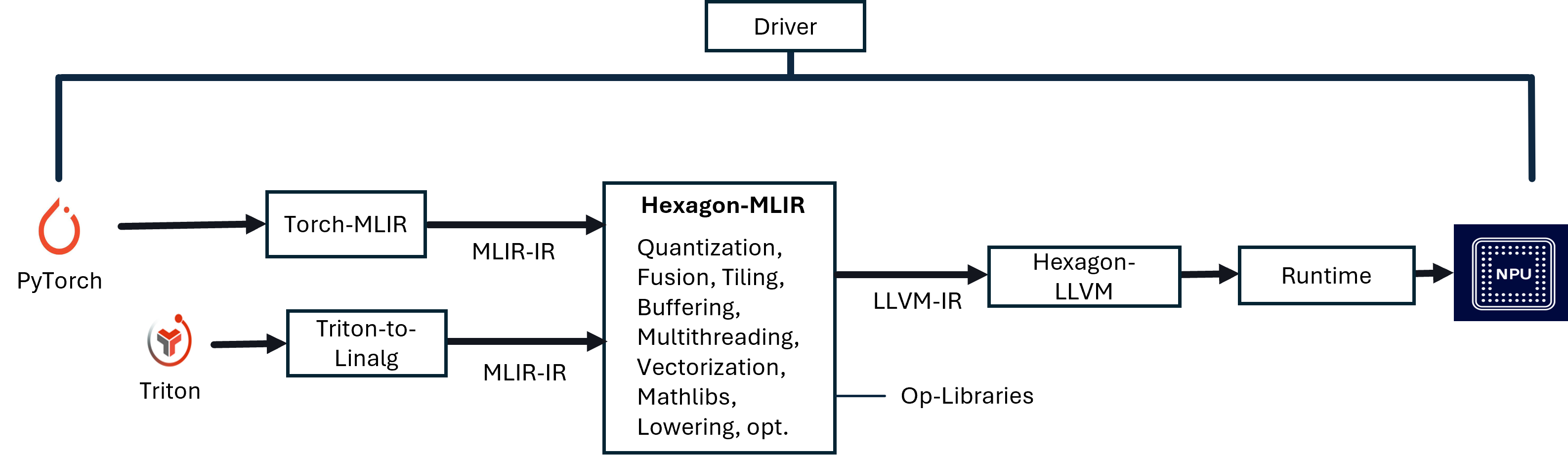}
\caption{Hexagon-MLIR — AI Compilation Stack}
\label{fig:hexagon_mlir_diagram}
\end{figure*}

\section{Introduction}

Hexagon-MLIR is a stack for compiling and executing Triton kernels \cite{TRITON} and PyTorch \cite{PyTorch} graphs on the
Qualcomm Hexagon Neural Processing Unit (NPU) \cite{Qualcomm2024NPU}.
Parts of the stack were open-sourced recently \cite{HexagonMLIRGitHub, HexagonMLIRBlog}.
Built using the MLIR framework \cite{mlir}, our compiler applies a structured sequence of conversion,
optimization, and lowering passes to exploit NPU architectural features — including Hardware Vector eXtension (HVX); 
multi-threading that exploits hardware threads and multiple HVX contexts;
memory hierarchy including Tightly Coupled Memory (TCM); interleaving of DMA transfers and computation;
optimized and vectorized math libraries; and a high-throughput matrix multiplication engine — to accelerate AI workloads.
 Hexagon-MLIR adopts a Generative Approach: by treating fusion as a first-class compiler pass,
we can generate specialized kernels for arbitrary long operator chains. 

The traditional approach to ML compilation often relied on \textit{operator libraries} which deliver peak performance but require
time-consuming, careful manual implementation \cite{TVM} \cite{Glow}.
With new operators and optimizations being written in Triton and other DSLs \cite{HALIDE, TensorComprehension} continuously,
a quick automated path from, e.g. PyTorch operators,
hand-written or Inductor-generated Triton kernels, to reasonably optimized binaries provides a significant deployment speed advantage.
Moreover, \textit{operator libraries} act as opaque barriers to optimization.
This fragmentation forces data to round-trip to DRAM between calls, creating a bandwidth bottleneck.
Much like region compilation \cite{Fisher, WMHwu}, which offered locality and parallelism exploitation in the 1990s, Triton
kernels via fusion \cite{neptune_fusion} provide opportunities for locality and parallelism exploitation that operator libraries can miss.
Hand-written assembly (peak-performance code) is faster for single operations, but automatic code generation is the only scalable solution 
for the rapid evolution of AI models.
Moreover, the Triton ecosystem continues to evolve with higher-level abstractions such as Helion \cite{Helion},
 which provides a PyTorch-centric DSL that compiles to Triton, further reducing the barrier to kernel development.
Hexagon-MLIR provides the architectural foundation to ingest these large regions (e.g. via PyTorch Inductor \cite{PyTorch, Ansel})
and compile them into binaries.

The challenge of efficiently compiling new large language models (LLMs) and recent Mixture-of-Experts architectures \cite{DeepSeekV3} 
remains an active area of research. Recent works have demonstrated that Triton can achieve state-of-the-art
performance for critical LLM components such as attention mechanisms \cite{TritonAttention, Attention, FlashDecoding}. 
Additionally, novel activation functions such as polynomial composition activations (PolyCom) \cite{PolyCom} have been proposed
to enhance the expressivity and dynamics of transformers, further motivating the need for flexible compilation frameworks
 that can efficiently support emerging architectural innovations. With the PyTorch 2.0 Inductor path, Triton kernels can be generated
  automatically. Alternatively, new operators or algorithms can be quickly and efficiently written in the Triton language.
The hand-optimized \textit{operator library} path is therefore not scalable.

Our compiler is built using MLIR \cite{mlir}. MLIR is not a specific compiler but a framework with which compilers can be constructed. 
One such AI compiler built using MLIR is IREE \cite{iree}, which provides portability via a Hardware Abstraction Layer (HAL). 
Another MLIR-based compiler is TPP (Tensor Processing Primitives) \cite{tpp}, which targets Intel and AMD processors and, according to the authors,
"achieves 90\% of the performance of hand-optimized equivalent programs." 
Compiler-based approaches have also been applied to distributed training optimization, as demonstrated by DeepCompile \cite{DeepCompile},
 which uses profiling-guided optimization passes to improve communication-computation overlap in distributed settings.
While wrapping hand-optimized libraries (e.g., Intel's approach with libxsmm) guarantees peak performance for standard operations, 
it fails to address the long tail of fusion patterns found in modern Transformers, and faces a combinatorial explosion of required kernels.

\section{Organization}
The remainder of this paper is organized as follows. Section~3 provides an overview of the Hexagon-MLIR compilation flow,
 formalizing key concepts including semantic preservation and the pass pipeline architecture. Section~4 presents detailed technical descriptions 
 of our core compiler passes—including Triton-to-Linalg conversion, the \texttt{linalg.generic} abstraction, operator fusion, tiling for memory hierarchy,
  multi-threading, double buffering, math library integration, and additional optimizations. We use motivating kernel examples (softmax and GELU) 
  throughout to illustrate the progressive transformation of code through the compilation pipeline. Section~5 presents experimental results and 
  performance analysis, demonstrating the effectiveness of individual optimizations (vectorization, multi-threading, double buffering) and their interactions
   as kernels transition between compute-bound and memory-bound regimes. Section~6 discusses related work in deep learning compilers and optimization techniques.
    Finally, Section~7 concludes with a summary of our contributions and directions for future work.

\section{Hexagon-MLIR: Overview}
Figure~\ref{fig:hexagon_mlir_diagram} provides an overview of the compilation flow. 
PyTorch models convert to Linalg and other MLIR core ops using Torch-MLIR \cite{TORCH_MLIR}. Similarly, Triton kernels convert to MLIR using  triton-to-linalg \cite{TRITON_SHARED}. 

\paragraph{Domains.}
Let $\M$ be the set of PyTorch models, $\K$ the set of Triton kernels, $\LIR$ the set of MLIR Linalg ops. Then we have:
\[
\begin{aligned}
f &: \M \longrightarrow \LIR \quad \text{(Torch-MLIR conversion)} \\
g &: \K \longrightarrow \LIR \quad \text{(Triton-to-Linalg conversion)}
\end{aligned}
\]
\paragraph{Semantics preservation.}
For all $m \in \M$ and $k \in \K$, lowering paths to Linalg should preserve the denotational semantics.
\[ \sem{m} = \sem{f(m)} \qquad\text{and}\qquad \sem{k} = \sem{g(k)}\]
While certain precision-related choices, e.g.  quantization, can  weaken strict semantic preservation, machine learning models are generally tolerant of such variations. Also, if custom named operations appear in the MLIR graph, we translate them to semantically equivalent MLIR core operations.

\paragraph{Hexagon-MLIR Pass Pipeline.}
The compilation process can be modeled as a sequence of morphisms between intermediate representations (IRs), where each pass transforms one IR into another:
\[
\begin{aligned}
\text{Obj}(\mathbf{IR}) &\;=\; \{\mathrm{IR}_{\alpha} \mid \alpha \text{ is an IR abstraction level}\} \\
\text{Hom}(\mathbf{IR}) &\;=\; \{p : \mathrm{IR}_{\alpha} \mapsto \mathrm{IR}_{\beta} \mid p \in \text{pass-pipeline}\}
\end{aligned}
\]
Passes could be classified into at least two types:
\begin{itemize}
    \item \textbf{Transformation:} rewrites the IR to simplify (e.g. cse, canonicalization) or to improve performance. E.g. vectorization pass:
    \[  p : \mathrm{IR}_{scalar} \mapsto \mathrm{IR}_{vectorized} \]     
    \item \textbf{Lowering:} lowers abstraction towards the target. E.g. structure-control-flow to unstructured control flow:
    \[  p : \mathrm{IR}_{\text{scf}} \mapsto \mathrm{IR}_{\text{cf}} \] 
\end{itemize}
Hexagon-MLIR pass pipeline has many passes but in this paper we focus on some main ones
\[
\begin{array}{|c|l|}
\hline
\textbf{Symbol} & \textbf{Description} \\
\hline
C  & \text{Canonicalization, CSE, or CP} \\
F  & \text{Operator Fusion} \\
L & \text{Layout transformation and propagation} \\
M & \text{Multi-threading} \\
p_{i} & \text{an optimization or lowering pass}\\
T  & \text{Tiling for memory hierarchy} \\
Q  & \text{Quantization} \\
V  & \text{Vectorization and related passes} \\
\hline
\end{array}
\]
The Hexagon-MLIR pipeline is a composition of passes applied on the input IR:
\[
\Pi = p_n \circ \dots \circ T \circ F \circ \dots \circ p_0 :
\mathrm{IR}_0 \longrightarrow \mathrm{IR}_{\text{final}}.
\]
The rest of the paper goes through in details on some of the important passes.

\section{Hexagon-MLIR: Details}
In this section we will deep-dive into some of the passes. We use a couple of examples to facilitate the explanation.

\subsection{Triton to Linalg}
\paragraph{Softmax} function is a widely used function in machine learning. Mathematically, for a given $\mathbf{z} = (z_1,\dots,z_K)$, softmax is:
\[
\sigma(z_i) = \frac{\exp(z_i - \max_j z_j)}{\sum_{k=1}^K \exp(z_k - \max_j z_j)}.
\]
Subtracting by $\max_j z_j$ improves numerical stability without changing the result.
\begin{lstlisting}[caption={Extract from Triton Softmax Kernel}, label={lst:SoftmaxTriton1}]
@triton.jit
def softmax_kernel(output_ptr, input_ptr, 
  ...
  row_minus_max = row - tl.max(row, axis=0)
  numerator = tl.exp(row_minus_max)
  denominator = tl.sum(numerator, axis=0)
  softmax_output = numerator / denominator
  ...
\end{lstlisting}
\paragraph{Triton Kernel.}
In Listing~\ref{lst:SoftmaxTriton1}, we show an excerpt of a Triton kernel that implements a numerically stable row-wise softmax. The kernel subtracts the maximum value in each row from all elements to prevent overflow during exponentiation. Then it computes $\texttt{tl.exp}$ and aggregates using a row-wise sum $\texttt{tl.sum}$ to form the normalization factor. Finally, the softmax probabilities are obtained by dividing each exponential by this normalizing factor.

\begin{figure*}[!t]
\centering
\begin{lstlisting}[caption={Linalg IR of the Softmax excerpt.},
                    label={lst:LinalgSoftmax}]
  %9 = linalg.generic 
         {indexing_maps = [affine_map<(d0) -> (d0)>, affine_map<(d0) -> (d0)>,
          affine_map<(d0) -> (d0)>],
          iterator_types = ["parallel"]}
         ins(%3, %8 : tensor<16384xf32>, tensor<16384xf32>)
         outs(%4 : tensor<16384xf32>) {
      ^bb0(%in: f32, %in_5: f32, %out: f32):
         %17 = arith.subf %in, %in_5 : f32
         linalg.yield %17 : f32
  } -> tensor<16384xf32>
  %10 = tensor.empty() : tensor<16384xf32>
  %11 = linalg.generic
          {indexing_maps = [affine_map<(d0) -> (d0)>, affine_map<(d0) -> (d0)>],
           iterator_types = ["parallel"]}
          ins(%9 : tensor<16384xf32>) outs(%10 : tensor<16384xf32>) {
      ^bb0(%in: f32, %out: f32):
         %17 = math.exp %in : f32
         linalg.yield %17 : f32
  } -> tensor<16384xf32>
\end{lstlisting}
\end{figure*}

The Triton kernel in Listing~\ref{lst:SoftmaxTriton1} is lowered by the Triton compiler to Triton-IR and then to Linalg by the Triton-to-Linalg converter. Listing~\ref{lst:LinalgSoftmax} shows the IR. It shows a sequence of \texttt{linalg.generic}s. We take a small detour here to explain \texttt{linalg.generic} as it forms an important part of the compilation flow.

\subsection{Linalg-Generic}
The \texttt{linalg.generic} is a key operator in the Linalg (Linear Algebra) dialect of MLIR. As an op it has a number of constructs - indexing map, iterator type, ins and outs, and linalg body.

\paragraph{Indexing maps:} Let $\mathcal{I} \subseteq \mathbb{Z}^d$ denote the \emph{iteration domain}, an integer lattice subject to affine constraints. A \texttt{linalg.generic} operation specifies a finite collection of affine maps
\[ \phi_\ell : \mathcal{I} \longrightarrow \mathbb{Z}^{n_\ell}, \]
each of which defines a view into the coordinate space of an input or output tensor $X_\ell$. Collectively, these maps determine how points in the common iteration domain are projected onto the indexing sets of the respective operands.
\paragraph{Iterator types:} Iteration domains labeled as \texttt{parallel} or \texttt{reduction}, encode interpretation of the computation along each axis in the body of the generic. The linalg body must be consistent with this declaration.    
\paragraph{Payload region:} The \texttt{linalg.body} which is a basic-block operating on scalars and yielding scalar results (outs) from scalar inputs. The body specifies the computation to be done at a single point in the iteration space of that \texttt{linalg.generic}. Each scalar is extracted from tensor operands using the indexing map and passed as block argument to the \texttt{linalg.body}.

The \texttt{linalg.generic} is a structured op that internalizes the geometry of iteration (index set, affine map, and iterator type) while delegating the computation to the linalg body. This design separates control over data traversal from computation semantics on the data, enabling principled transformations prior to lowering to explicit loops. For instance, every named op (e.g., \texttt{linalg.add}, \texttt{linalg.conv\_2d})  uniquely lowers to \texttt{linalg.generic}. This is key to its usefulness which ensures uniform transformation, optimization and lowering machinery across the sea of ops in ML. Although you can have linalg.generic operating on buffers, the best practice is to use \texttt{linalg-on-tensors}, which we do, as it provides the benefits of SSA representation. 

It must be mentioned on balance that many ops that have characteristics such as -- shape-inconsistent outputs; data-dependent flow; non-affine maps etc -- cannot be represented by a single \texttt{linalg.generic}. Important ML ops such as \texttt{softmax}, \texttt{prefix-sum}, \texttt{TopK} etc need other control flow ops along with \texttt{linalg.generic} to make them representable. Also, it must be mentioned that an intermediate abstraction layer was added to Linalg in the form of \texttt{linalg.contraction}, \texttt{linalg.elementwise}. These also lower uniquely to \texttt{linalg.generic}.

\subsection{Operator Fusion}
\begin{figure*}[!t]
\centering
\begin{lstlisting}[caption={IR of Softmax excerpt after Fusion},
                   label={lst:FusedLinalg}]
  ...
  %7 = linalg.generic
           {indexing_maps = [affine_map<(d0) -> (d0)>,
                             affine_map<(d0) -> (d0)>],
            iterator_types = ["parallel"]}
        ins(%3 : tensor<16384xf32>) outs(%4 : tensor<16384xf32>) {
      ^bb0(%in: f32, %out: f32):
         %11 = arith.subf %in, %extracted : f32
         %12 = math.exp %11 : f32
         linalg.yield %12 : f32
     } -> tensor<16384xf32>
  ...
\end{lstlisting}
\end{figure*}
\textbf{Operator fusion} is a semantics-preserving transformation that combines multiple structured operations (e.g., \texttt{linalg.generic}) into a single operation, eliminating intermediate materialization and improving data locality.

Consider two operations:
\[
X \xrightarrow{\;P\;} Y \xrightarrow{\;Q\;} Z,
\]
where \(P\) and \(Q\) are \texttt{linalg.generic} operations with iteration domains \(\mathcal{I}_P\) and \(\mathcal{I}_Q\).
The \textbf{producer} \(P\) writes elements of \(Y\) using:
\[  \phi^{P}_{\text{out}} : \mathcal{I}_P \to \mathrm{dom}(Y), \]
while the \textbf{consumer} \(Q\) reads elements of \(Y\) using:
\[ \phi^{Q}_{\text{in}} : \mathcal{I}_Q \to \mathrm{dom}(Y).  \]

A case of fusion combines the two \texttt{linalg.generic} into one so that instead of writing back the entire tensor \(Y\), in the body of the fused generic scalar value \(x\) of \(X\) is read  to produce scalar value \(y\) of \(Y\) that is immediately consumed to produce \(z\) of \(Z\). Therefore, if
\[
\begin{aligned}
P &= \texttt{linalg.generic}\big\{\mathcal{I}_P,\; \{\phi^P_\ell\},\;
\mathbf{y} \coloneqq f(\mathbf{x})\big\}, \\
Q &= \texttt{linalg.generic}\big\{\mathcal{I}_Q,\; \{\phi^Q_\ell\},\;
\mathbf{z} \coloneqq g(\mathbf{y})\big\},
\end{aligned}
\]
then the fused operation is:
\[
\texttt{linalg.generic}\big\{\mathcal{I},\; \{\tilde{\phi_\ell}\},\;
\mathbf{z} \coloneqq g\big(f(x)\big)\big\}.
\]
Listing~\ref{lst:FusedLinalg} shows the fused \texttt{subf} and \texttt{exp} generics from Listing~\ref{lst:SoftmaxTriton1}. The example in Listing~\ref{lst:FusedLinalg} is rather simple (identity-map) for illustration but in general fusion has to consider - broadcast, transposes, and re-computations.

\subsection{Tiling for Memory Hierarchy}
\begin{figure*}[!t]
\centering
\begin{lstlisting}[caption={GELU after Fusion and Tiling},
                   label={lst:TiledGELU}]
#map = affine_map<(d0) -> (d0)>
module attributes {llvm.target_triple = "hexagon"} {
func.func @gelu_kernel(%arg0: memref<*xf32> {tt.divisibility = 16 : i32}, ..) {  
  %cst_2 = arith.constant 7.978850e-01 : f32  
  ..
  %4 = scf.for %arg8 = %c0 to %c1048576 step %c262144
        iter_args(%arg9 = %0) -> (tensor<1048576xf32>) {
      %extracted_slice = tensor.extract_slice %3[%arg8] [262144] [1]
           : tensor<1048576xf32> to tensor<262144xf32>
      %5 = bufferization.alloc_tensor() 
             copy(%extracted_slice) {memory_space = 1 : i64} : tensor<262144xf32>  
      ..
      %7 = linalg.generic
             {indexing_maps = [#map, #map], iterator_types = ["parallel"]}
             ins(%5 : tensor<262144xf32>) outs(%6 : tensor<262144xf32>) {
      ^bb0(%in: f32, %out: f32):        
        ..
        %13 = arith.mulf %12, %cst : f32
        %14 = math.exp %13 : f32
        %15 = arith.addf %14, %cst_0 : f32
        %16 = arith.subf %cst_0, %14 : f32
        %17 = arith.divf %16, %15 : f32
        ..
        linalg.yield %20 : f32
      } -> tensor<262144xf32>
      %inserted_slice = tensor.insert_slice %7 into %arg9[%arg8] [262144] [1]
           : tensor<262144xf32> into tensor<1048576xf32>
      scf.yield %inserted_slice : tensor<1048576xf32>
    } {all_parallel, tiled_generic}
    ..
  }
} 
\end{lstlisting}
\end{figure*}

Tiling partitions large tensors that typically reside on the slower but large shared memory (DDR) into smaller \texttt{tiles} that are copied onto the smaller but faster local Tightly Coupled Memory (TCM). Computations such as vectorized HVX operations are then done on data that is on TCM. High reuse of data that is brought or generated on TCM is important to offset the latency of transfer latency between DDR and TCM. Operator fusion that we saw in the previous section helps improve vector register level reuse. Reuse across fused operators is facilitated be retaining tensor results in TCM.

Let us look at the tiling mechanism in Hexagon-MLIR in more details. Given a \texttt{linalg.generic} with iteration domain $\mathcal{I} \subseteq \mathbb{Z}^d$, a tiling  vector $\mathbf{t} = (t_1, \dots, t_d)$, and an optional interchange vector, tiling introduces two levels of iteration: an outer loop consisting of \texttt{scf::for} or \texttt{scf::forall}, and the body of the loop remains a \texttt{linalg.generic} that now operates on a tile on TCM. 

To exploit memory hierarchy, the tiling pass in Hexagon-MLIR introduces two distinct memory spaces: DDR and TCM. Data transfers between DDR and TCM are performed via DMA operations. Tiling is applied at the \texttt{Linalg-on-Tensor} level, where the pass inserts explicit data movement operations using \texttt{bufferization.alloc\_tensor} with memory-space attributes. These attributes enable the compiler to later generate DMA transfers between DDR and TCM. Furthermore, the tiling pass incorporates loop interchange to position vectorizable dimensions innermost and annotates tiled loops to facilitate subsequent optimizations such as software pipelining.
We will come back to softmax but for now let us turn to another interesting kernel for now - Gaussian Eror Linear Unit (GELU), to explore tiling, vectorization and double buffering.
\[\text{GELU}(x) \approx 0.5x \left(1 + \tanh\left(\sqrt{\frac{2}{\pi}}(x + 0.044715x^3)\right)\right)
\]

The GELU kernel after fusion and tiling is shown in Listing~\ref{lst:TiledGELU}. Bufferization has not happended yet so the tiled generic still operates on tensors. Notice the \texttt{scf.for} loop, where in each iteration a slice of the larger input tensor is calculated using  via \texttt{tensor.extract\_slice}. That slice is then copied from the default memory space to TCM. The body of the \texttt{linalg.generic} operation computes on the data brought into TCM. The generic's body although  presently looks like scalar code, after further transformations it will become multi-threaded HVX operations. Next, we will multi-thread this tiled generic.

\subsection{Multi-threading}
\begin{figure*}[!t]
\centering
\begin{lstlisting} [caption={GELU after async-threads creation},
                    label={lst:GeluMT}]
  ...                  
  %12 = async.create_group %11 : !async.group
  scf.for %arg9 = %c0_10 to %c262144_11 step %c8_12 {
    %token = async.execute {
      %subview_14 = memref.subview %2[%arg9] [8] [1] 
              : memref<262144xf32, 1> to memref<8xf32, strided<[1], offset: ?>, 1>
      ..
      scf.for %arg10 = %c0 to %c8 step %c1 {
        %14 = memref.load %subview_14[%arg10] : memref<8xf32, strided<[1], offset: ?>, 1>
        %15 = arith.mulf %14, %cst_1 fastmath<fast> : f32
        ...
        %21 = math.exp %20 fastmath<fast> : f32
        %22 = arith.addf %21, %cst_0 fastmath<fast> : f32
        %23 = arith.subf %cst_0, %21 fastmath<fast> : f32
        %24 = arith.divf %23, %22 fastmath<fast> : f32
        ...
        memref.store %27, %subview_15[%arg10] : memref<8xf32, strided<[1], offset: ?>, 1>
      }
      async.yield
    }
    %13 = async.add_to_group %token, %12 : !async.token
  }
  async.await_all %12
  ...  
\end{lstlisting}
\end{figure*}
Qualcomm NPUs have hardware multi-threading to support parallel execution of \texttt{Hexagon Vector eXtensions (HVX)} instructions. A vector context is a set of resources that enables independent execution of HVX instructions, including a vector register file and a vector predicate file. Multiple hardware threads can run in parallel, each utilizing a different vector context. In this section we will show multi-threading of the GELU kernel we tiled in previous section. We have two multi-threading schemes. One uses \texttt{scf::for} starting point and the other uses \texttt{linalg.generic}. We will show the latter one here. Our multi-threading uses MLIR's Async dialect which provides a structured fork–join IR:
\begin{itemize}
    \item \texttt{async.execute} creates an async region, returning a token (and optionally values) with \texttt{async.yield}.
    \item \texttt{async.await} waits on a token/value by suspending the current co-routine.
    \item \texttt{async.create\_group}, \texttt{async.add\_to\_group}, and \texttt{async.await\_all} implement barriers over multiple tokens.
\end{itemize}

The asynchronous representation lowers to LLVM coroutines (\texttt{llvm.coro.id}, \texttt{llvm.coro.begin}, \texttt{llvm.coro.suspend}, \texttt{llvm.coro.end}) together with the MLIR async runtime.
Hexagon-MLIR exploits HVX multi-threading with a two-stage pipeline:
\begin{enumerate}
    \item \emph{Form-Virtual-Threads}: partitions work and encodes parallel structure using \texttt{scf.forall}. 
    \item \emph{Form-Async-Threads}: lowers that structure to MLIR's Async dialect. This approach keeps parallel semantics declarative and analyzable.
\end{enumerate}

\paragraph{Stage-1:} Analyzes linalg.generic to decide if parallelization is profitable using a polytope size heuristic.
It performs a tiling over parallel iterators to produce scf.forall (virtual threads) using block distribution when ranges divide evenly; using  block-cyclic for thread size balancing.

\paragraph{Stage-2:} Each \texttt{scf.forall} is rewritten into an explicit asynchronous fork–join pattern. First, \texttt{async.create\_group(N)} establishes a barrier for \texttt{N} threads. A surrounding \texttt{scf.for} then iterates over the tiles, and each iteration spawns:
\begin{quote}
\texttt{async.execute \{ /* body of the forall tile */ \} $\rightarrow$ token}
\end{quote}

The resulting token is added to the group using:
\begin{quote}
\texttt{async.add\_to\_group(group, token)}
\end{quote}
Finally, \texttt{async.await\_all(group)} enforces completion before any dependent work proceeds. This approach preserves the original parallel semantics while preparing for coroutine-based code generation.

Listing~\ref{lst:GeluMT} shows the IR of theGELU kernel after multi-threading. Its input IR was the tiled GELU kernel of Listing~\ref{lst:TiledGELU}.

\subsection{Double Buffering}

\begin{figure*}[!t]
\centering
\begin{lstlisting} [caption={GELU after double-buffering},
                    label={lst:GeluDB}] 
// -----// IR Dump After hexagon-double-buffer-generic-s2   // -------------- //
func.func @gelu_kernel(%arg0: memref<*xf32> {tt.divisibility = 16 : i32}, ..) {
  %false = arith.constant false
  // allocate ping-pong buffers on TCM
  %ping_buffer = memref.alloc() {alignment = 2048 : i64} : memref<262144xf32, 1>
  %pong_buffer = memref.alloc() {alignment = 2048 : i64} : memref<262144xf32, 1>
  %dma_ping_tag = memref.alloc() : memref<1xi32> 
  %dma_pong_tag = memref.alloc() : memref<1xi32>                          
  ... 
  /// Double Buffering - Prologue
  scf.if %loop_executes_at_least_once {
    %subview = memref.subview %Input[0] [262144] [1] 
                 : memref<1048576xf32, ..>> to memref<262144xf32,..>>
    memref.dma_start %subview[%c0], %ping_buffer[%c0],%c262144,%dma_ping_tag[%c0]
          : memref<262144xf32,.>>, memref<262144xf32,1>, memref<1xi32>
    ...
  } {db_generic = 0 : i64, db_prologue}
  /// Double Buffering - Kernel (main loop)
  scf.for %arg8 = %c0 to %c1048576 step %c262144 {
    %is_ping_stage = memref.load %ping_pong_toggle[] : memref<i1>
    ...
    scf.if %is_ping_stage {    
      scf.if %is_not_last_iteration {
        // Prefetch to pong buffers for next iteration.
        %subview_17 = memref.subview %Input[%4] [262144] [1]
                          : memref<1048576xf32, ..>> to memref<262144xf32, ..>>
        memref.dma_start %subview_17[%c0], %pong_buffer[%c0],
                 %c262144, %dma_pong_tag[%c0]
              : memref<262144xf32, ..>>, memref<262144xf32, 1>, memref<1xi32>
        ...
      } {db_prefetch}
      memref.dma_wait %dma_ping_tag[%c0], %c262144 : memref<1xi32>
      ...
      // Do multi-threaded HVX execution on ping buffers on TCM 
      scf.forall (%arg9) = (0) to (262144) step (65536) {
        %subview_17 = memref.subview %ping_buffer[%arg9] [65536] [1] 
                       : memref<262144xf32, 1> to memref<65536xf32, ...>, 1>
        ...
        scf.for %arg10 = %c0 to %c65536 step %c32 {
          %subview_19 = memref.subview %subview_17[%arg10] [32] [1]
             : memref<65536xf32,..>, 1> to memref<32xf32, ..>, 1>
          ...    
          %15 = math.exp %14 fastmath<fast> : vector<32xf32>
          ...
          vector.transfer_write ...  : vector<32xf32>, memref<32xf32, ..>, 1>
        }
      }  
      memref.dma_start ... // write-back ping buffer results.
      memref.store %false, %ping_pong_toggle[] : memref<i1> // ping-pong state
    } {db_ping_kernel}
    scf.if %is_pong_stage {
      ... // pong-stage
    } {db_pong_kernel}  
    ... // deallocations etc.
}  
\end{lstlisting}
\end{figure*}

Double buffering in software pipelining is a technique of overlapping computation with data movement, thereby mitigating memory latency in hierarchical memory systems. This approach leverages two buffers (ping and pong) — while one is actively engaged in computation the other is concurrently being filled (and then the roles swap) - enabling data transfers and computations to happen in parallel.

In AI/ML workloads, where tensor operations dominate and data sizes often exceed the capacity of fast local memory, double buffering becomes critical. The interleaving of communication and computation transforms memory-bound kernels towards latency-tolerant pipelines, exploiting parallelism across memory and compute units. Similar overlapping techniques have been explored in distributed AI systems, where compiler-based approaches enable native overlapping optimizations for distributed workloads \cite{TRITON_DISTRIBUTED}.

The implementation in hexaogn-mlir introduces double buffering in two layered passes that separate structural software pipelining from asynchronous data movement.

\paragraph{Stage 1: Structural Transformation}
Stage~1 finds canonical single-buffered, tiled loops  \texttt{scf.for} that are in ``normal form''. The loops have attributes indicating they originate from tiled \texttt{linalg.generic}. The tiling leaves per-tile schedule triplets 
\[
\texttt{memref.subview} \rightarrow \texttt{memref.alloc} \rightarrow \texttt{memref.copy},
\]
along with the compute region and the write-back sequence. Double buffering pass synthesizes a ping-pong flow: a guarded prologue that prefetches the first tile into the ping buffers, and a rebuilt kernel with two mutually exclusive sub-kernels (annotated \texttt{db\_ping\_kernel}/\texttt{db\_pong\_kernel}). Each sub-kernel issues a prefetch for the next iteration, clones the compute with \texttt{alloc} remapped to the current ping/pong buffers, and reconstructs the stores by re-materializing subviews at the current induction variable. A memory-resident boolean \texttt{toggle} drives ping$\leftrightarrow$pong selection and is updated each iteration, while attributes like \texttt{db\_generic} (unique ID) and \texttt{db\_prologue}/\texttt{db\_prefetch} provide IR anchors for downstream passes. This keeps the transformation deterministic and hazard-free, preserving the original tiling semantics and establishing the overlap between communication (prefetch) and computation without yet committing to a specific asynchronous transport.

\paragraph{Stage 2: Asynchronous DMA Integration}
Stage~2 materializes true asynchronous transfers and synchronization. It parses the Stage~1 IR using the injected attributes to recover the prologue, the kernel, and the ping/pong sub-schedules. . The pass then rewrites all preload and prefetch \texttt{memref.copy} operations into \texttt{memref.dma\_start}---allocating distinct ping/pong tags so that each phase can be waited on independently---and inserts \texttt{memref.dma\_wait} immediately after the corresponding prefetch blocks to ensure data is resident before compute. For the store-back path, it similarly replaces each \texttt{memref.copy} with \texttt{dma\_start} followed by \texttt{dma\_wait} (using separately allocated store tags), and emits balanced deallocations for all tags after the kernel. This design decouples legality/scheduling (Stage~1) from transport semantics (Stage~2), enabling latency hiding through explicit DMA start/wait without perturbing the compute IR.

Listing~\ref{lst:GeluDB} shows the IR after double-buffering. The IR is annotated in C++ style comments and some variables given custom names for easier comprehension.

We have been illustrating our flow with GELU IRs so lets recap here. Listing~\ref{lst:FusedLinalg} showed at an earlier stage, right after fusion. Fusion happens on \texttt{linalg-on-tensor}.Then we did tiling to bring slices from DDR to TCM in 
Listing~\ref{lst:TiledGELU}. The tiled code forms bridge between tensors and memrefs (buffers).  Some other passes such as bufferization run subsequently. Then we showed multi-threading for HVX in the IR in Listing~\ref{lst:GeluMT}. After that comes double buffering which we showed in this section. There is still more lowering that happens after this leading finally to LLVM-IR with runtime calls.

\subsection{Math Library}
In AI/ML workloads, replacing transcendental functions like \texttt{exp} with low-degree polynomial approximations is a common way to trade controlled numerical error for gains in throughput. We invoke Qualcomm Hexagon Library (QHL) -- a set of math libraries pre-optimized for vectorized computations.

MLIR also provides \texttt{polynomial approximation} expansions of math operations for fast approximations that do not rely on any custom library functions. The general approach e.g. to approximate $e^x$, firstly express it as 
\[
e^x = e^{\epsilon + b} = e^{\epsilon + n \log 2} = e^\epsilon \cdot 2^n.
\]
Then, as $\epsilon$ is chosen to be small, Taylor series up to some terms can be used to compute $e^{\epsilon}$. Horner's formula is often used to reduce the number of multiplications and additions to compute the polynomial.

In addition to above, there are other optimizations and algorithms used to produce fast implementations of math operations. One  example is \textbf{fast inverse square root}.

\subsection{Optimizations and Lowering}
The flow  discussed in previous section leads to successful compilation of PyTorch and Triton kernels by our tool. We did not cover all the passes and optimizations that are in place or in development. We also did not discuss MLIR core passes that we use e.g. bufferization, deallocation, and lowering to llvm as they are quite well known. We identify convolutions and matrix multiplications earlier on in our flow and generally they go to a dedicated engine for matrmix multiplication. The data needs to be laid out in a particular way and we use \texttt{linalg.pack/unpack} to do the data-space transformations. To reduce the number of packs/unpacks we use layout-propagation.

\section{Results and Analysis}

\begin{table*}[t]
  \centering
  \small
  \caption{Speedups by vectorization.}
  \label{tab:vectorization}
  \begin{tabularx}{\textwidth}{l c Y}
    \toprule
    Test & Speedup & Notes\\
    \midrule
    
    Gaussian Error Linear Unit (GELU)    & 63.9x  & float16,  elements = 16384 \\
    Gaussian Error Linear Unit (GELU)    & 16.1x  & float32,  elements = 16384 \\      
    Double Buffered GELU & 15.4x         & float32,  element= 16 x 16384, double buffered \\    
    Flash Attention              & 4.7x  & float32, N\_CTX=1024, DIM\_HEAD=64, BLOCK\_N=64 \\   
  
    Root Mean Square Norm (RMS-norm)     & 46.5x  & float16, elements = 127 x 513,  \\
    Sigmoid Linear Unit (SiLU)  & 4.8    & float16,  elements = 16384 \\
    Sigmoid Linear Unit (SiLU)  & 7.1    & float32,  elements = 16384 \\
    Vec-Add-2D                  & 40.6x  & float32,  elements = 64 x 16384 \\     
    \bottomrule
  \end{tabularx}
\end{table*}

The results in this section are organized to map the compiler optimizations described in previous sections to the performance improvements shown in the accompanying tables and bar charts. We enable individually and in combination set of passes in the Hexagon‑MLIR pipeline: fusion (F) removes intermediate materialization and improves locality; tiling (T) moves working sets into TCM; HVX vectorization (V) exploits wide SIMD execution; multi-threading (M) distributes tiled work across hardware HVX contexts; and double buffering (DB) overlaps DMA transfers with computation. This structured presentation allows each optimization described in the technical sections to be correlated with its observed runtime effect, and highlights how these passes interact—sometimes constructively, sometimes competitively—as kernels transition from compute‑bound to memory‑bound regimes.

\subsection{Vectorization Speedups}
Table~\ref{tab:vectorization} summarizes speedups from HVX vectorization across some kernels. GELU achieves $63.9\times$ on \texttt{float16} (16,384 elements) and $16.1\times$ on \texttt{float32}, indicating that HVX width utilization and data type packing are primary drivers of throughput. 

Gains from vectorization of RMS-norm reaches $46.5\times$ on  127$\times$513 shape, confirming that stencil linalg.elementwise dominated computations, even with some reductions, gain from vectorization. Just to recap, RMS-Norm is defined as:
\[
\text{RMSNorm}(x) = \frac{x}{\sqrt{\frac{1}{d} \sum_{i=1}^{d} x_i^2 + \epsilon}} \cdot g
\]
The computation part of its triton kernel is shown in Listing~\ref{lst:TritonRMSNormKernel}.
\begin{lstlisting}[caption={Extract from Triton RMS Norm Kernel},
       label={lst:TritonRMSNormKernel}]
sq = tl.sum(x * x, axis=0) / NUM_COLS
rms = tl.sqrt(sq + EPSILON)
y = (x / rms) * g
\end{lstlisting}

Table~\ref{tab:vectorization} also includes results for the \textbf{Sigmoid Linear Unit (SiLU)} activation. SiLU is defined as:
\[
\text{SiLU}(x) = x \cdot \sigma(x) = x \cdot \frac{1}{1 + e^{-x}},
\]
We get good $5-8$x gains. The computation is dominated by the  term $e^x$. For \texttt{float16}, performance decreases when operations are internally promoted to \texttt{float32}.

\subsection{Single vs. Multi-threading}
\begin{figure}[t]
\centering
\begin{tikzpicture}
  \begin{axis}[
    ybar,
    width=0.7\linewidth,
    height=5.5cm,
    scale only axis,
    xlabel={Problem Sizes},
    ylabel={Speedup},
    ymin=0,
    grid=both,
    grid style={dashed,gray!30},
    bar width=10pt,
    symbolic x coords={8192,16384,32768,65536,131072,262144,524288,1048576},
    xtick=data,
    xticklabel style={font=\footnotesize, rotate=60, anchor=east},
    xticklabels={8K,16K,32K,64K,128K,256K,512K,1M},
    tick label style={/pgf/number format/fixed},
    enlarge x limits=0.12,
    legend style={at={(0.02,0.98)}, anchor=north west, draw=none, fill=white},
  ]
    \addplot[
      draw=blue,
      fill=blue!30,
    ] coordinates {
      (8192,    0.581259151)
      (16384,   0.710769231)
      (32768,   2.27617801)
      (65536,   1.904201681)
      (131072,  2.381455399)
      (262144,  3.690217391)
      (524288,  3.948700866)
      (1048576, 3.401162791)
    };
    \addlegendentry{Speedup}
  \end{axis}
\end{tikzpicture}
\caption{Speedup of multi-threaded GELU over increasing problem-sizes.}
\label{fig:speedup_gelu_st_over_mt}
\end{figure}
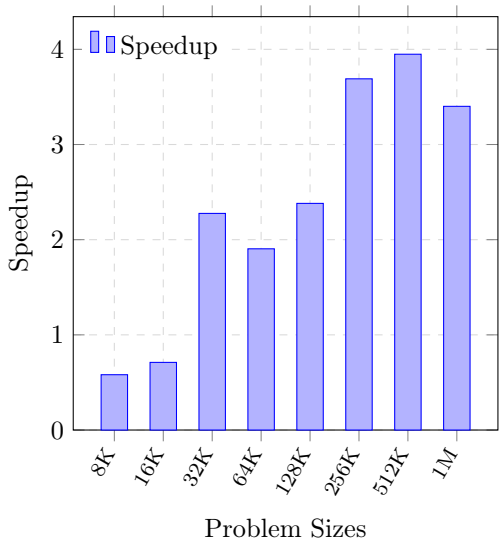

Figure~\ref{fig:st-mt-gelu-comparison} compares single-threaded execution against HVX multi-threading over HVX contexts across increasing problem sizes (8K $\to$ 1M elements). 
At small sizes ($\leq$ 16K), multi-threaded runtimes are higher due to thread start-up, token management, and barrier overheads. Beyond 32K elements, multi-threading consistently outperforms single-threaded execution with speedups ranging from $2.28\times$ (32K) to $3.95\times$ (512K), and $3.40\times$ at 1M elements (see the speedup bar chart). These gains emerge when the parallel payload amortizes the fixed costs of \texttt{async.execute} and \texttt{async.await\_all}. At 1M elements, speedup dips slightly compared to 512K, reflecting increasing pressure on TCM and DMA queues. In such case re-tuning could be beneficial. Figure~\ref{fig:speedup_gelu_st_over_mt} presents the speedup, defined as the ratio of single-threaded execution time to multi-threaded execution time, in a more interpretable format. 

\subsection{Double Buffering}
\begin{figure}[t]
\centering
\begin{tikzpicture}
  \begin{axis}[
    ybar,
    bar width=14pt,
    ymin=0,
    enlarge x limits=0.2,
    axis lines=box,
    grid=both,
    major grid style={dashed, gray!60},
    minor grid style={dotted, gray!40},
    minor tick num=4,
    ylabel={execution time},
    xtick=data,
    xticklabel style={font=\small, rotate=15, anchor=east},
    symbolic x coords={Scalar, Vec, Vec+MT, Vec+MT+DB},
    nodes near coords,
    nodes near coords align={vertical},
    minor tick num=1,
  ]
    \addplot [fill=blue!30!white, draw=black] 
    coordinates {(Scalar,128065) (Vec,3252) (Vec+MT,3000) (Vec+MT+DB,2700)};
  \end{axis}
\end{tikzpicture}
\caption{Vector-Add 2D performance across optimization passes.}
\label{fig:vec-add-2d-values}
\end{figure}
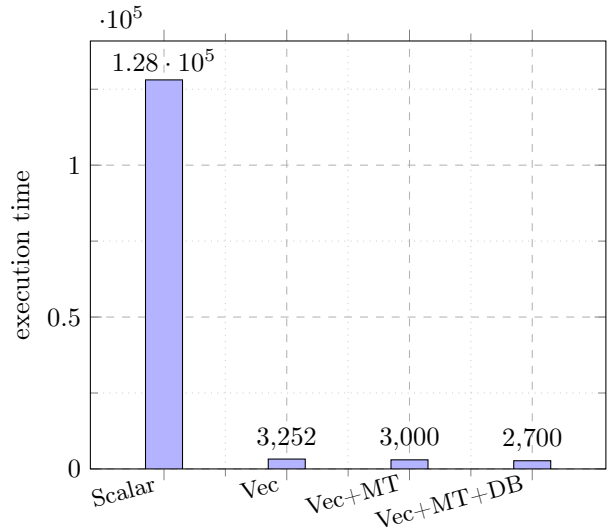

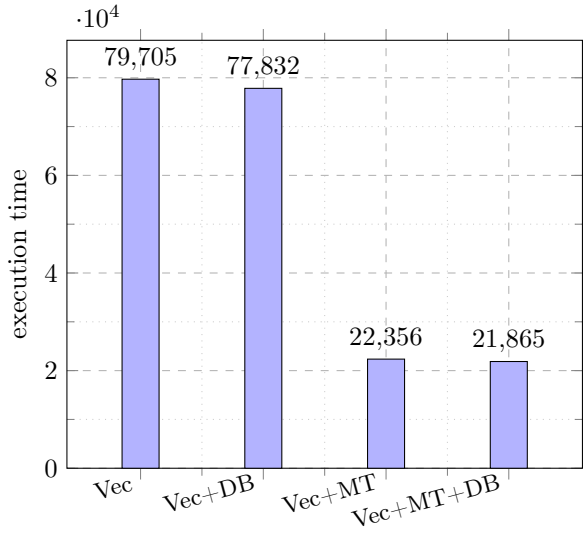
\begin{figure}[t]
\centering
\begin{tikzpicture}
  \begin{axis}[
    ybar,
    bar width=14pt,
    ymin=0,
    enlarge x limits=0.2,
    axis lines=box,
    grid=both,
    major grid style={dashed, gray!60},
    minor grid style={dotted, gray!40},
    minor tick num=4,
    ylabel={execution time},
    xtick=data,
    xticklabel style={font=\small, rotate=15, anchor=east},
    symbolic x coords={Vec, Vec+DB, Vec+MT, Vec+MT+DB},
    nodes near coords,
    nodes near coords align={vertical},
    minor tick num=1,
  ]
    \addplot [fill=blue!30!white, draw=black] 
    coordinates {(Vec,79705) (Vec+DB,77832) (Vec+MT,22356) (Vec+MT+DB,21865)};
  \end{axis}
\end{tikzpicture}
\caption{Multi-pass analysis of Exponent Series.}
\label{fig:multi_pass_exponent_series}
\end{figure}

\begin{figure}[t]
\centering
\begin{tikzpicture}
  \begin{axis}[
    ybar,
    bar width=14pt,
    ymin=0,
    enlarge x limits=0.2,
    axis lines=box,
    grid=both,
    major grid style={dashed, gray!60},
    minor grid style={dotted, gray!40},
    minor tick num=4,
    ylabel={execution time},
    xtick=data,
    xticklabel style={font=\small, rotate=15, anchor=east},
    symbolic x coords={Vec, Vec+DB, Vec+MT, Vec+MT+DB},
    nodes near coords,
    nodes near coords align={vertical},
    minor tick num=1,
  ]
    \addplot+[fill=blue!30!white, draw=black] 
    coordinates {(Vec,18505) (Vec+DB,16924) (Vec+MT,7059) (Vec+MT+DB,6221)};
  \end{axis}
\end{tikzpicture}
\caption{Multi-pass analysis of GELU.}
\label{fig:multi_pass_gelu_series}
\end{figure}
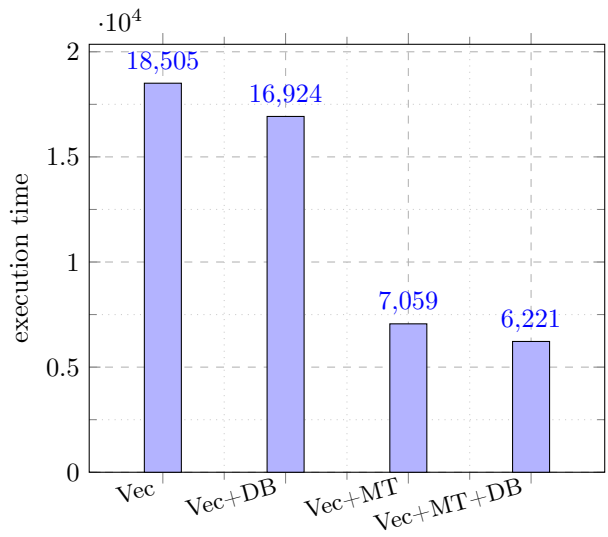

Figure~\ref{fig:vec-add-2d-values} shows improvements with vectorization, multi-threading (MT), and double-buffering (DB). Vectorization gives significant gains, but if we take it as a given (base case), then MT and DB improve performance further.  Figure~\ref{fig:multi_pass_gelu_series} shows the gains from double buffering and other optimizations for GELU. Similarly, Figure~\ref{fig:multi_pass_exponent_series} shows improvements for the compute-intensive exponent series kernel.

\subsection{Multi-pass Interaction}
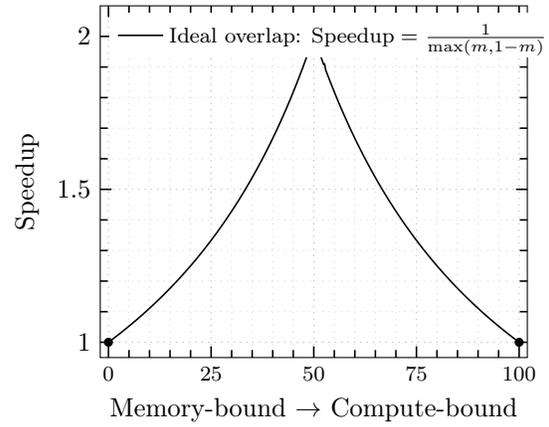
\begin{figure}[t]
\centering
\begin{tikzpicture}
\begin{axis}[
  width=0.9\linewidth,
  ylabel={Speedup},
  xlabel={Memory-bound \(\rightarrow\) Compute-bound},
  xmin=0, xmax=100,
  ymin=0.95, ymax=2.1,
  grid=both,
  grid style={dotted, gray!60},
  tick style={black, semithick},
  minor tick num=4,
  minor grid style={dotted, gray!30},
  xtick={0,25,50,75,100},
  x tick label style={font=\footnotesize},
  enlarge x limits=0.02,
  legend style={
    at={(0.02,0.98)},
    anchor=north west,
    draw=none,
    font=\footnotesize
  },
  legend cell align=left,
]
  \addplot[
    semithick,
    draw=black!70!black,
    samples=400,
    domain=0:100,
  ]
  {(x <= 50) ? 1/(1 - x/100) : 1/(x/100)};
  \addlegendentry{Ideal overlap: \(\text{Speedup} = \frac{1}{\max(m,1-m)}\)}

  \addplot[
    only marks,
    mark=*,
    mark size=1.5pt,
    draw=black!70!black,
  ] coordinates {(0,1) (50,2) (100,1)};

  \node[anchor=east, font=\footnotesize] at (axis cs:0,0.98) {Memory-bound};
  \node[anchor=west, font=\footnotesize] at (axis cs:100,0.98) {Compute-bound};
\end{axis}
\end{tikzpicture}
\caption{Speedup for double buffering with perfect overlap as memory fraction varies.}
\label{fig:speedup-mem-comp-curve}
\end{figure}

We now turn to multi-pass interactions. Techniques like vectorization and multi-threading mainly reduce computation time, which makes memory transfer overhead more pronounced. Double buffering helps by overlapping data movement with computation, boosting overall performance. However, the net effect can be that an optimized kernel becomes memory-bound.  Figure~\ref{fig:speedup-mem-comp-curve} illustrates an idealized case: when memory transfers completely dominate (100\%), double buffering offers no benefit; likewise, if computation dominates entirely, the technique is again ineffective. In practice, most workloads fall between these extremes and so double-buffering provides gain, and secondary effects—such as overheads and scheduling — play an important role. Applying optimization passes selectively provides additional insight into kernel characteristics and these secondary interactions.

\section{Related Work}
Li et al.~\cite{li2021deep_learning_compiler_survey} provide a comprehensive survey of deep learning (DL) compilers, analyzing their design architectures and optimization strategies. Although published five years ago, the concepts remain relevant today. The paper explains how DL compilers transform models from various frameworks into optimized code for heterogeneous hardware, focusing on multi-level intermediate representations (IR) and both frontend and backend optimizations. It discusses key techniques such as operator fusion, memory planning, layout transformation, hardware mapping, and auto-tuning.

Yifan et al \cite{neptune_fusion}, present Neptune, a tensor compiler that supports loop fusion algorithms for reduction operators, that satisfies data dependency by recomputation and automatically derives required algebraic repairs. Two instances, Rolling Update Fusion and Split-K Fusion, are particularly suited for optimizing attention-like operators. They implemented Neptune on top of the Apache TVM schedule tensor compiler and the Triton tile tensor compiler. A more recent paper \cite{dl_compiler_strategy} by the same author on \textit{Strategies for Graph Optimization in Deep Learning Compilers}  delves into front-end optimization techniques, which according to the author are instrumental in refining the structure of computational graphs, thereby significantly bolstering the efficiency of neural network training and inference phases.

Chin et al, in the paper \cite{HLOEnv} \textit{HloEnv: A Graph Rewrite Environment for Deep Learning Compiler Optimization Research}, analyze the impact on the performance of an optimization pass/pipeline on the HLO dataset from two perspectives: the proportion of the dataset affected by that pass, and the average change in performance as a result of that pass (runtime ratio w/ and w/o the pass). 

Optimizing both computation speed and memory size and transfer requirements is important for an AI compilation package. An XLA compiler extension that adjusts the computational data-flow representation of an algorithm according to a user-specified memory limit is discussed in \cite{arutaev2022xsafe}.Benoit et al. \cite{MODEL_memopt} present MODeL, an ILP approach that optimizes the lifetime and memory location of the tensors used to train neural networks to reduce the memory usage of models.

\section{Conclusion}       
This work presented Hexagon-MLIR, an MLIR-based compilation stack designed  for Qualcomm's Hexagon NPU for AI workloads. By combining MLIR's extensible infrastructure and hardware-aware passes, Hexagon-MLIR delivers a systematic lowering pipeline for PyTorch models and Triton kernels. Key optimizations—such as operator fusion, tiling, HVX vectorization, multi-threaded execution, and double buffering—enable efficient utilization of Tightly Coupled Memory (TCM), asynchronous DMA transfers, and specialized math libraries. Experimental results demonstrate  performance gains across representative kernels, validating the effectiveness of our approach.
Beyond performance, this work illustrates how MLIR's design principles—structured IRs, declarative transformations, and staged lowering—provide a strong foundation for building compilers targeting AI workloads. While the current implementation achieves promising results, ongoing efforts focus on expanding coverage for emerging models, refining scheduling heuristics, and improve feature support and performance. Ultimately, Hexagon-MLIR aims to deliver a scalable, high-performance solution for deploying next-generation AI workloads on Hexagon NPUs.

% Figures
\begin{figure}[t]
\centering
\begin{tikzpicture}
  \begin{axis}[
      width=0.9\linewidth,
      xmode=log,
      log basis x=2,
      xmin=8192, xmax=1048576,
      xtick={8192,16384,32768,65536,131072,262144,524288,1048576},
      xticklabel style={/pgf/number format/fixed},
      xlabel={Problem Size (num-elements)},
      ylabel={Execution Time (\si{normalized})},
      grid=both,
      minor tick num=4,
      minor grid style={dotted, gray!30},
      grid style={dotted, gray!60},
      tick style={black, semithick},
      legend style={
           at={(0.02,0.98)},
           anchor=north west,
           draw=none,
           font=\footnotesize
        },
      legend cell align=left,
    ]
    \addplot+[semithick, mark=o, mark size=2.5pt, color={rgb,255:red,31; green,119; blue,180}]
      coordinates {
        (8192, 397)
        (16384, 462)
        (32768, 1739)
        (65536, 1133)
        (131072, 2029)
        (262144, 7469)
        (524288, 11854)
        (1048576, 14625)
      };
    \addlegendentry{Single-threaded}    
    \addplot+[semithick, mark=square*, mark size=2.3pt, color={rgb,255:red,214; green,39; blue,40}]
      coordinates {
        (8192, 683)
        (16384, 650)
        (32768, 764)
        (65536, 595)
        (131072, 852)
        (262144, 2024)
        (524288, 3002)
        (1048576, 4300)
      };
    \addlegendentry{Multi-threaded}
   \end{axis}
 \end{tikzpicture}
 \caption{Execution times of single-threaded vs multi-threaded GELU kernel implementations.}
 \label{fig:st-mt-gelu-comparison}
\end{figure}
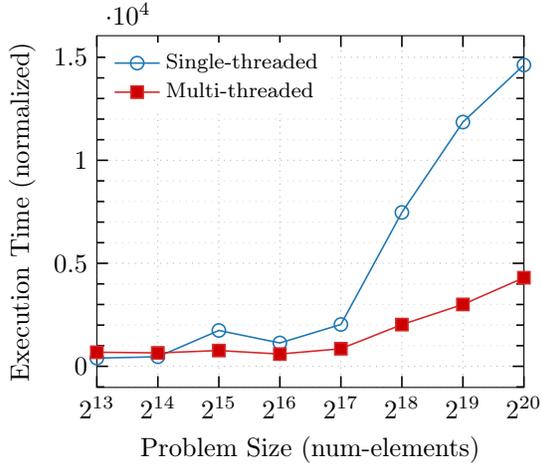

\bibliographystyle{plain}
\bibliography{reference_arxiv}

\end{document}